# MAGNETIC IRREVERSIBILITY IN ULTRAFINE $ZnFe_2O_4$ PARTICES

G.F. Goya[†], H.R. Rechenberg

*Instituto de Física, Universidade de São Paulo, CP 66318, 05315-970 São Paulo, Brazil.*

M. Chen and W. B. Yelon

*University of Missouri Research Reactor, Columbia, Missouri 65211, USA.*

## Abstract

Pure ultrafine $ZnFe_2O_4$ particles have been obtained from mechanosynthesis of the ZnO and $Fe_2O_3$ oxides. The average grain diameter was estimated from x-ray diffraction to be $<d> = 36(6)$ nm. Refinement of neutron diffraction (ND) data showed that the resulting cubic spinel structure is oxygen-deficient, with ~7% of $Fe^{3+}$ ions occupying the tetrahedral A sites. Magnetization curves taken at 4.2 K showed absence of saturation up to fields H = 9 Tesla, associated to a spin-canted produced by the milling process. Field-cooled (FC) and zero-field-cooled (ZFC) curves showed irreversible behavior extending well above room temperature, which is associated to spin disorder. Annealing samples at 300 °C yields an average grain size $<d> = 50(6)$ nm, and ~16% of $Fe^{3+}$ ions at A sites. Partial oxygen recovery is also deduced from neutron data refinement in annealed samples. Concurrently, decrease of magnetic irreversibility is noticed, assigned to partial recovery of the collinear spin structure. Complex Mössbauer spectra were observed at room temperature and 80 K, with broad hyperfine field distributions spanning from ~10 T to ~40 T. At T = 4.2 K, hyperfine field distributions indicate high disorder in Fe local environments. The above data suggest the existence of Fe-rich clusters, yielding strong superexchange interactions between Fe ions at A and B sites of the spinel structure.

[†] Corresponding author. Electronic mail: *goya@unizar.es*

## Introduction

The theoretical and technological lowest size limits of magnetically ordered systems are still an open question[1,2], of major relevance to applications of nanostructured systems in high-density magnetic recording devices. Ultrafine particles of many ferrites have been successfully obtained by numerous chemical routes, such as reverse micelle synthesis, co-precipitation, thermal decomposition and aerogel process. High-energy ball milling (HEBM) has been used as an alternative route to obtain novel materials through solid-state reactions, since mechanochemical reactions usually yield highly metastable phases.

The structure of the spinel oxides $AB_2O_4$ consists of a close-packed FCC arrangement of oxygen atoms, with two nonequivalent crystallographic sites (A and B) in the structure[3]. For $ZnFe_2O_4$, the $Zn^{2+}$ and $Fe^{3+}$ distribution at A and B sites within the structure can be represented by the formula $[Zn_\delta Fe_{1-\delta}]^A[Zn_{1-\delta}Fe_{1+\delta}]^BO_4$, where $\delta$ is the inversion parameter. Values from $\delta=1$ (normal) to $\delta=0.21$ are reported in the literature, depending on the synthesis method[4-6]. For samples with $\delta=1$, antiferromagnetic order of the Fe ions appears below $T_N = 9\text{-}11$ K[4,7,8]. This low ordering temperature compared to other $MFe_2O_4$ spinels with magnetic M (for which $T_C \sim 500\text{-}800$ K[3]) is due to the weak $J_{B-B}$ superexchange interactions between Fe sites, in contrast to the $J_{A-B}$ stronger interactions.

## Experimental Procedure

The $ZnFe_2O_4$ samples were obtained by mechanosynthesis from the precursor ZnO and $\alpha\text{-}Fe_2O_3$ oxides, adding acetone as a liquid carrier. Powders were milled during 1320 h with the vials closed, and afterwards heated at 70 °C until complete drying (sample ZA). A fraction of this mixture was further annealed in air at 300 °C for 1 h (sample ZH). Samples were examined by x-ray diffraction (XRD) using Cu-K$\alpha$ radiation. Room temperature neutron diffraction patterns were taken at the University of Missouri Research Reactor (MURR), scanning within the 5 º ≤ 2Θ ≤ 105 º range at λ = 1.487 Å neutron wavelength. Mössbauer measurements were performed between 4.2 and 296 K, using a conventional constant-acceleration spectrometer in transmission geometry with a $^{57}$Co source in a Rh matrix. Isomer shifts are referred relative to α-Fe at room temperature. Field cooled (FC) and Zero field cooled (ZFC) magnetization curves were performed within the 4.2 K < T < 300 K range using H = 10G, in a commercial SQUID magnetometer.

## Experimental Results

Neutron data taken at room temperature for *as prepared* and annealed samples are shown in Figure 1. For both the *as milled* (ZA) and annealed (ZH) samples, the ND patterns could be indexed as cubic $ZnFe_2O_4$ spinel (Fd3m space group) phase, with cell volumes $V_{cell}$ = 596.0(1) Å$^3$ and 596.33(5) Å$^3$, respectively. An important sloping background in the pattern of sample ZA can be noticed from figure 1, which disappears for the annealed sample. This scattering could be originated from a fraction of the liquid carrier strongly bound to the particles in the *as prepared* samples, which is eliminated after heating at 300 °C. The refined parameters indicate oxygen deficiency in the *as prepared* sample, which decreases after heat treatment, in agreement with the reducing condition inside the closed containers during milling. The obtained Fe and Zn populations at A and B sites in the structure yield an inversion degree δ = 0.93 and 0.84 for ZA and ZH samples, respectively. This means that the amount of $Fe^{3+}$ ions at the tetrahedral A sites increases from 7% to 16% after annealing. The average grain sizes <d> estimated from the x-ray patterns, by applying the Scherrer equation on the three strongest lines, were <d> = 36(6) nm and <d> = 50(6) nm for ZA and ZH samples, respectively.

ZFC-FC magnetization curves are shown in figure 2. It is observed that irreversible behavior extends well above room temperature. In addition, there is a broad maximum centered at $T \sim 220$ K, which might be related to the collective blocking of the particles. M(H) curves taken at $T = 4.2$ K for ZA (not shown) indicated magnetic irreversibility below $H \sim 6$ T, which is lost after the annealing process (sample ZH). Concurrently, the effective magnetic moment extracted from the M(H) curves increases from $1.79\mu_B$/F.U. for ZA to $3.48\mu_B$/F.U. for ZH sample.

Room temperature Mössbauer spectra (figure 3) of both samples showed a central doublet, with a marked background curvature. These spectra were fitted using a distribution of quadrupolar P($\Delta$) plus a magnetic P(H) hyperfine fields. For sample ZA, the P($\Delta$) distributions showed three components extending up to $\approx 1.5$ mm/s, whereas for ZH sample the components are narrower and there is no contribution above $\approx 1.0$ mm/s. Both hyperfine field distributions show a maximum near 10 T, extending up to $\sim 40$ T. The fitted isomer shift values span from 0.28(1) mm/s to 0.33(1)mm/s, indicating that iron is in a $Fe^{3+}$ state. At $T = 80$ K, a magnetic signal develops and the central doublet is no longer observable, although weak relaxation effects are still noticed at this temperature in both samples. The maximum of the hyperfine field distribution is shifted to $H\sim 48$ T, with smaller contributions extending down to $\sim 30$ T. At $T= 4.2$ K, where relaxation effects are absent, the spectra showed only a magnetic sextet with broad lines (see figure 4) due to unresolved magnetic hyperfine fields. The hyperfine field distributions used to fit the spectra showed two main peaks, located at $H \approx 52$ T and $\approx 50$ T, and a third component at $H \approx 48$ T, amounting $\sim 15$ % of the total spectral area. At $T=4.2$K, no relaxation effects are present and a six-line magnetic pattern is observed in both samples. However, the observed hyperfine field distributions indicate the presence of different Fe environments. The oxygen deficiency observed from ND data suggests that superexchange paths between magnetic Fe atoms might be not fully connected, resulting in different magnetic fields at each Fe site. At a given octahedral site, a magnetic Fe ion may have different numbers of $Fe^{3+}$ and $Zn^{2+}$ as its nearest tetrahedral sites. The resulting superexchange interaction of each Fe at a B site with the six A nearest-neighbor cations (which in turn determines the hyperfine field value) will depend on the number of $Zn^{2+}$ at the nearest A sites. Additionally, the electron density at the Fe nucleus in B site will also depend on the number of Fe and Zn nearest neighbors, giving different isomer shifts. The resulting distribution of exchange interactions and electronic densities at different sites, caused by a deviation of the ideal normal structure, will generate a distribution of hyperfine fields.

## Discussion

It has been previously reported that ball milling of ferrites yielded large changes of inversion degrees as compared to the same crystalline phase[4,6]. In the present work, where complete mechanosynthesis of the spinel phase has been achieved, ND data showed that the final deviation from the normal configuration in the *as prepared* sample is only 7%, increasing to a 16.4% for ZH sample. In the framework of a simple collinear antiferromagnetic (Néel-type) order below $T_N$, the increase in magnetic moment might be proportional to the amount of unpaired Fe moments at the A sublattice. The expected increase in magnetic moment from ND data would be from $\mu_{eff} = 0.45\mu_B/FU$ for ZA to $\mu_{eff} = 0.80\mu_B/FU$ for ZH samples. This is significantly less than the observed increase of the magnetic moment at 4.2 K after annealing, and thus other processes are needed to account for it. Originally proposed by Coey[9] to explain Mössbauer data from $\gamma$-$Fe_2O_3$ nanoparticles, it was later established that the decrease of saturation magnetization in ferro- or ferrimagnetic nanoparticles is due to spin canting of the original collinear structure[2,10,11]. However, two different models have been proposed to explain the fundamental origin of the spin canting, related to surface and finite-size effects respectively[2,10-13]. Whereas in the former model the disordered spins at the particle surface generate a shell-core structure, in the latter the spin structure is altered by strong anisotropy effects. In the present $ZnFe_2O_4$ nanoparticles, the increase of the effective moment at 4.2 K for ZH sample is concurrent with the observed partial oxygen recovery, suggesting that the collinear spin structure might be recovered after annealing.

Early neutron diffraction studies on $ZnFe_2O_4$[5,8] have indicated that the spin arrangements below $T_N$ are far more complex than the collinear Néel-type AFM ordering. More recently, using ND and muon spin rotation[4,7] it was found that, whereas long range magnetic order appears at $T\sim T_N$, short-range AF correlations are present in a wide range of temperature ($0.5T_N < T < 8T_N$). It was proposed[4] that these short-range correlations were originated in small ($\sim$1-3 nm) regions with superantiferromagnetic behavior. Due to the fast relaxation rate of these small AF 'clusters' (in the GHz range), Mössbauer spectra will appear paramagnetic above $T_N$. Since $ZnFe_2O_4$ is paramagnetic at room temperature, the relaxation effects observed in Mössbauer spectra at 296 K cannot be attributed to superparamagnetic effects in ZA and ZH samples. In addition, the high disorder in Fe hyperfine fields observed at 4.2 K indicates the presence of very dissimilar Fe coordination. The above data suggests that Fe-rich clusters might result from HEBM process. Having the same cubic spinel structure, the high population of Fe at A and B sites within these clusters could lead to magnetic behavior that resembles the $Fe_3O_4$ phase, yielding the observed relaxation at room temperature. Additional neutron diffraction measurements at low temperatures in these mechanosynthesized samples could help to clarify this point.

ACKNOWLEDGEMENTS: We are grateful to Dr. J.Z. Jiang for fruitful discussions. This work was partially supported by the Fundação de Amparo à Pesquisa do Estado de São Paulo (FAPESP).

**Figure Captions:**

Figure 1. Neutron diffraction patterns for *as prepared* (ZA) and annealed (ZH) $ZnFe_2O_4$, showing the observed (points), calculated (solid line) and difference patterns (lower curve).

Figure 2. ZFC-FC curves for *as prepared* (circles) and annealed (triangles) samples.

Figure 3. Room temperature Mössbauer spectra for *as prepared* (ZA) and annealed (ZH) samples. Solid lines are the best fits using quadrupolar ($\Delta$) and hyperfine field (H) distributions. The resulting P($\Delta$) and P(H) distributions are shown in the insets.

Figure 4. Mössbauer spectra at T=4.2 K for *as prepared* (ZA) and annealed (ZH) samples. Solid lines are the best fits using a distribution of hyperfine fields. The resulting distributions P(H) are shown in the insets.

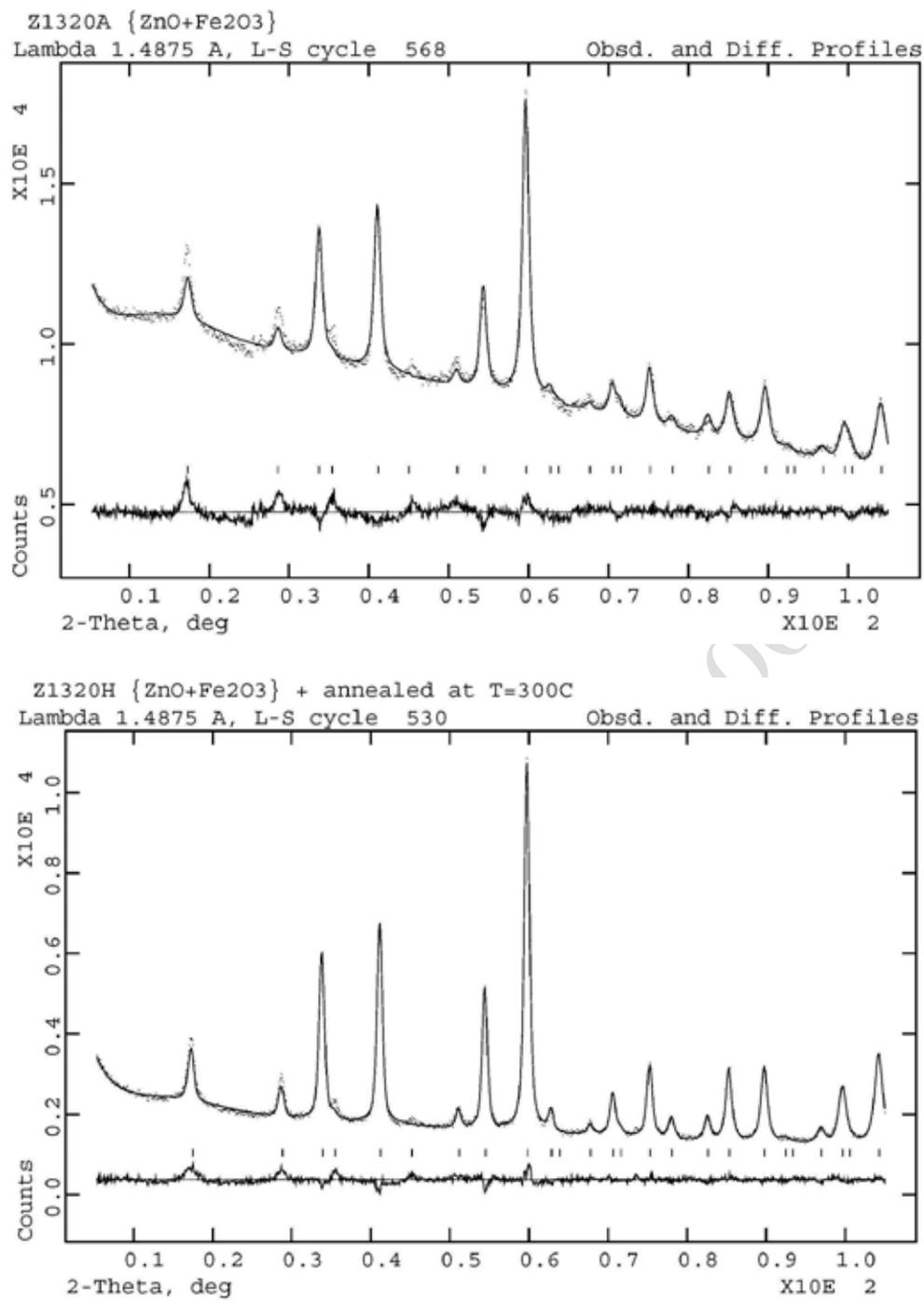

FIGURE 1

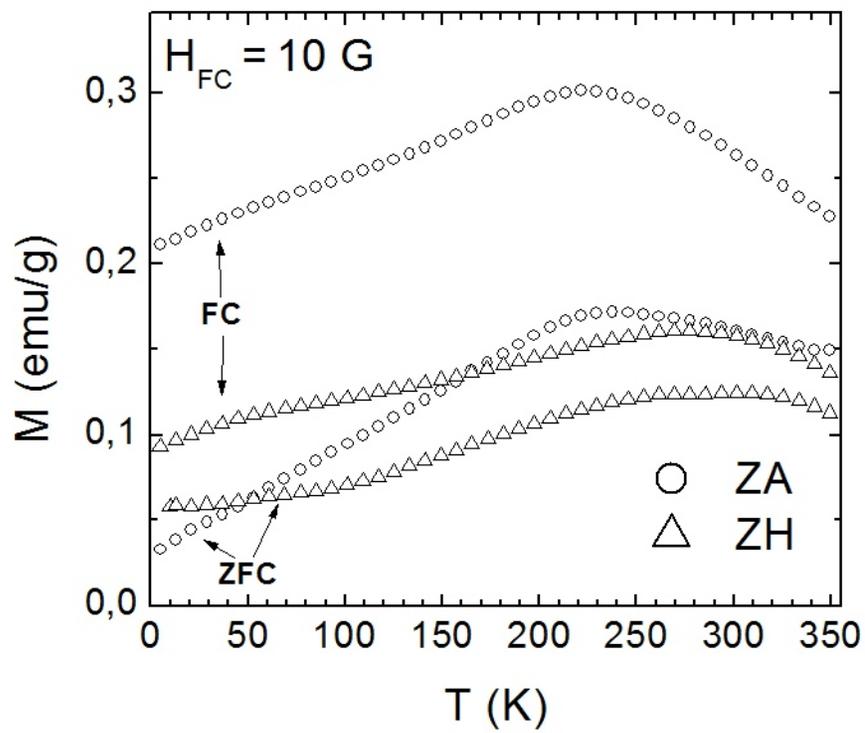

FIGURE 2

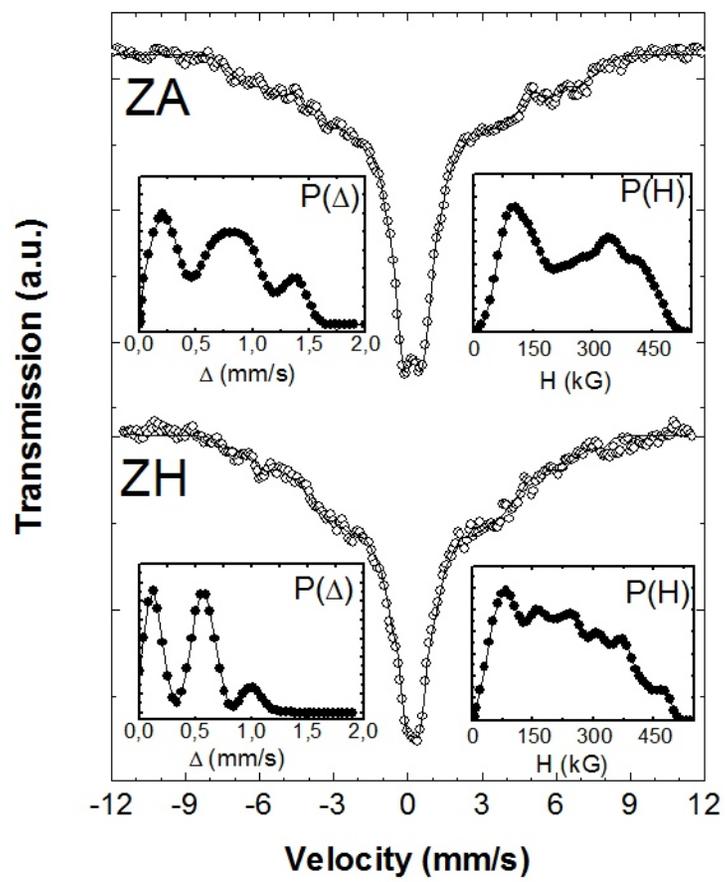

FIGURE 3

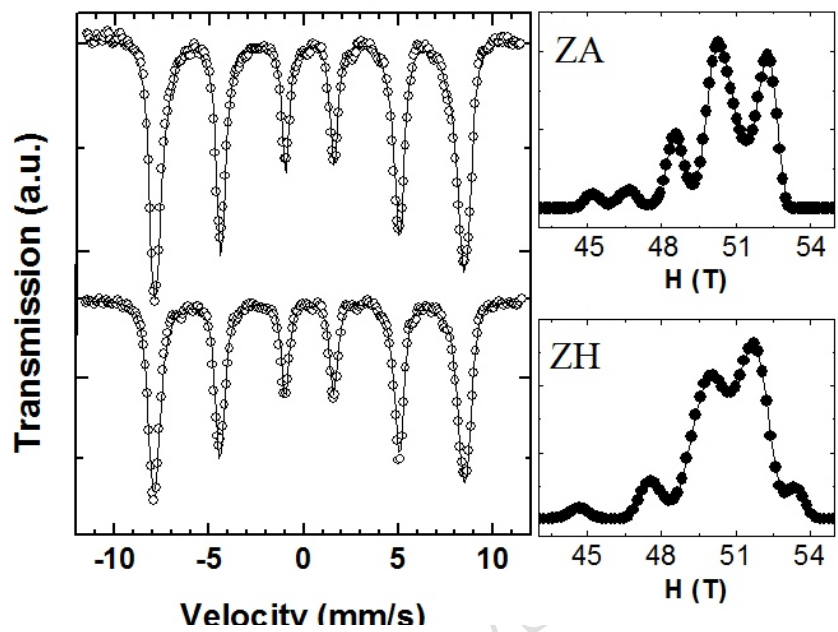

FIGURE 4